\documentclass{jfm}
\usepackage{graphicx}
\usepackage{natbib}

\ifCUPmtlplainloaded \else
   \checkfont{eurm10}
   \iffontfound
     \IfFileExists{upmath.sty}
       {\typeout{^^JFound AMS Euler Roman fonts on the system,
                    using the 'upmath' package.^^J}%
        \usepackage{upmath}}
       {\typeout{^^JFound AMS Euler Roman fonts on the system, but you
                    dont seem to have the}%
        \typeout{'upmath' package installed. JFM.cls can take advantage
                  of these fonts,^^Jif you use 'upmath' package.^^J}%
       }
   \else
   \fi
\fi

\ifCUPmtlplainloaded \else
   \checkfont{msam10}
   \iffontfound
     \IfFileExists{amssymb.sty}
       {\typeout{^^JFound AMS Symbol fonts on the system, using the
                 'amssymb' package.^^J}%
        \usepackage{amssymb}%

       }{}
   \fi
\fi

\ifCUPmtlplainloaded \else
   \IfFileExists{amsbsy.sty}
      {\typeout{^^JFound the 'amsbsy' package on the system, using it.^^J}%
      \usepackage{amsbsy}}
     {}
\fi

\newcommand\ie{\mbox{\textit{i.e.\ }}}

\newcommand\Rey{\mbox{\textit{Re}}}   
\newcommand\I{\mbox{\textit{P}}}     
\newcommand\Oh{\mbox{\textit{Oh}}}   
\newcommand\We{\mbox{\textit{We}}}   
\newcommand\Rmax{\mbox{\textit R}_{\rm max}}        
\newcommand\Vret{\mbox{\textit V}_{\rm ret}}        
\newcommand\thetaR{\theta_{\rm R}}        
\newcommand\RI{\mbox{\textit R}_{\rm I}}        
\newcommand\VI{\mbox{\textit V}_{\rm I}}        
\newcommand\tauv{\tau_{\rm v}}        
\newcommand\taui{\tau_{\rm i}}        

\title[Drop Retraction upon impact]{Retraction 
dynamics of aquous drops upon impact on 
nonwetting surfaces.}

\author[Denis Bartolo, Christophe Josserand and Daniel Bonn]%
{DENIS BARTOLO$^1$
CHRISTOPHE JOSSERAND$^2$
and DANIEL BONN$^{1,3}$
}
\affiliation{$^1$Laboratoire de Physique Statistique de l'ENS, 24 Rue Lhomond, 
75231 Paris cedex 05, France\\[\affilskip]
$^2$Laboratoire de Mod\'elisation en M\'ecanique, 
CNRS-UMR 7606, Case 162, 4 place Jussieu, 75252 
Paris C\'edex 05-France\\[\affilskip]
$^3$ van der Waals-Zeeman Institute, University 
of Amsterdam, Valckenierstraat 65, 1018 XE 
Amsterdam, The Netherlands}

\pubyear{2005}
\volume{xxx}
\pagerange{xxx--yyy}
\begin{document}
\date \today
\maketitle

\begin{abstract}
We study the impact and subsequent retraction dynamics of liquid droplets upon 
high-speed impact on hydrophobic surfaces. Performing extensive experiments, we show that 
the drop retraction rate is a material constant and does not depend on the impact velocity. We 
show that when increasing the Ohnesorge number, $\Oh=\eta/\sqrt{\rho R_{\rm I} 
\gamma}$, the retraction, \ie 
dewetting, dynamics crosses over from a capillaro-inertial regime to a capillaro-viscous 
regime. We rationalize the experimental observations by a simple but robust 
semi-quantitative model for the solid-liquid contact line dynamics inspired by the standard 
theories for thin film dewetting.
\end{abstract}
\section{Introduction: Drop Impact on Solid Surfaces}
\begin{figure}
\centerline{\includegraphics[width=12cm]{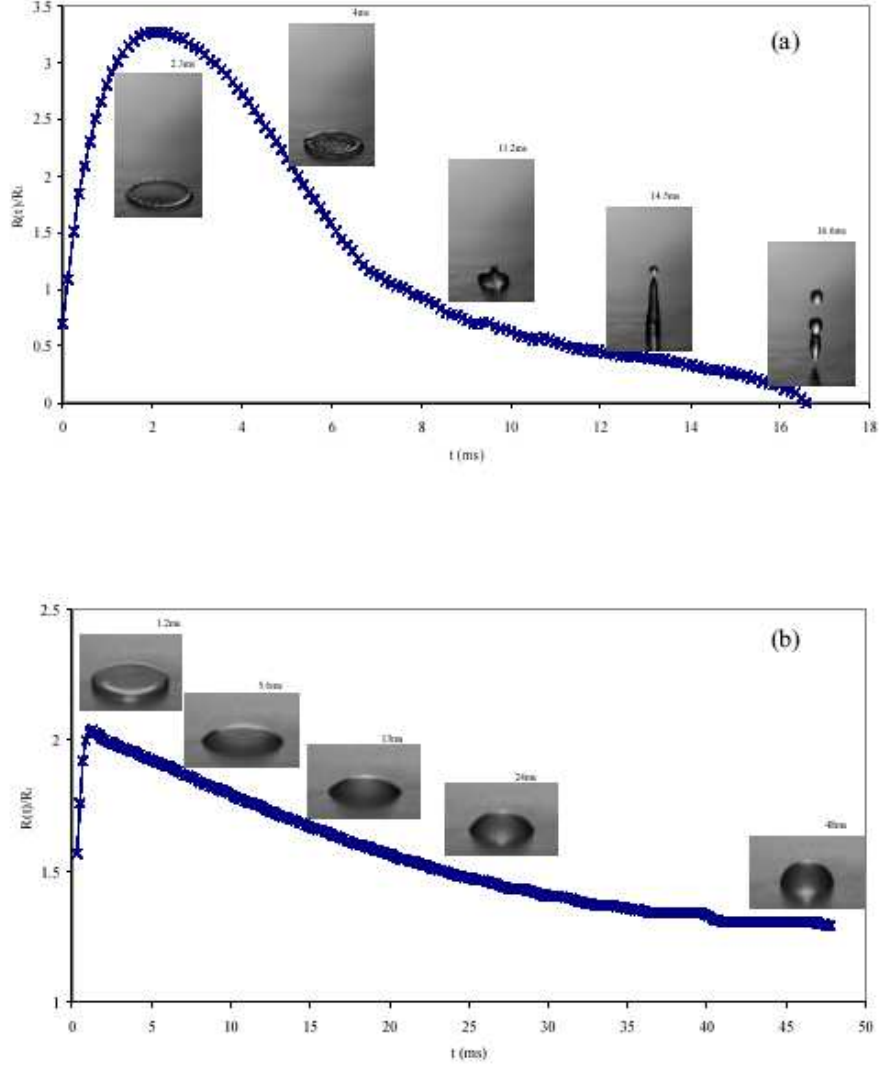}}
\caption{Temporal evolution of the contact radius of droplets upon impact and retraction. The
radii are normalized by those of the spherical droplets before impact. The pictures show the shape of
the droplets at the different stages of retraction.  Droplet radius is $1$ mm, impact speed is 
$2 \;{\rm m \cdot s^{-1}}$: a) pure water, b) viscous water-glycerol mixture, viscosity $50\; {\rm mPa\cdot s}$. }\label{fig:Rwater}
\end{figure}
Drops impacting onto solid surfaces are important for a large number of applications: for instance,
almost all spray coating and deposition processes rely ultimately on the interaction of a droplet 
with a surface. A large variety of phenomena can be present during drop impacts, from splashes to 
spreading, and from large wave surface deformation to rebound (see \cite[]{Rein93} and 
references therein). 

Research on 
drop impacts has a long history, starting with the pioneering studies of Worthington and later on with the
famous photographs of Edgerton\cite[]{Worth,Edge54}.  Most of the previous work on drop impact focused 
on determining the 
maximum diameter a drop is capable of covering upon impact   \cite[]{fukai93,Roisman2002,Clanet2004}. 
However, the practical problem of 
deposition can be very different if one wants to efficiently deposit some material on the surface. 
This is especially grave when the surface is not wetted by the liquid, as is illustrated by the high-speed
video pictures in Fig.\ref{fig:Rwater} for the impact of a water droplet. It can 
be observed that the drop expands rapidly, due to the large speed with which it arrives at the 
surface. However, due to the hydrophobicity of the surface, subsequently the drop retracts 
violently, leading to the ejection of part of the droplet from the surface: we observe 
droplet rebound. It is this "rebound" that is the limiting factor for deposition in many 
applications, for instance for the deposition of pesticide solutions on hydrophobic 
plant leaves \cite[]{Bergeron}. We study here the impact and subsequent retraction of aqueous drops onto a 
hydrophobic surface, and seek to understand the dynamics of expansion and retraction of the droplets. 

In general, these problems are difficult because 
for most practical and laboratory situations, 
three forces play an important role: the capillarity and 
viscous forces, and the inertia of the droplets. 
We try and disentangle the effects of the three 
forces here by performing systematic experiments, varying both the importance of
viscous and inertial forces.

We provide 
experimental evidence for the existence of two 
distinct retraction regimes. In both regimes, capillary forces are the motor behind the droplet retraction, and are, for the first regime countered by inertial forces. In the second regime the main force slowing down the retraction is viscous. We also show that, perhaps surprisingly, 
the drop retraction rate (the retraction speed 
divided by the maximum radius) does not depend on the impact 
velocity for strong  enough impacts. The dimensionless number 
that governs the retraction rate is found to be 
the Ohnesorge number,  $\Oh=\eta/\sqrt{\rho R_{\rm I} 
\gamma}$, with $\eta$ the viscosity, $\rho$ the 
liquid density,
$R_{\rm I}$ the impacting drop radius, and $\gamma$ the surface 
tension. The Ohnesorge number therefore compares 
the dissipative (viscous) forces to the non-dissipative (capillary and inertial) forces. The crossover 
between the two regimes is found to happen at a critical 
Ohnesorge number on the order of $0.05$ . 

In order to develop a better
understanding for the different regimes that are 
encountered, particularly the retraction dynamics in these
regimes, we propose two simple hydrodynamic 
models inspired by the standard description of 
thin film dewetting dynamics. These simple models provide a 
simple but quite robust picture that allows us to
rationalize the retraction rate in both regimes.

In order to be able to say something about the speed of retraction, one also needs to understand the maximum radius to which the droplet expands. Combining our results with those obtained by  \cite[]{Clanet2004} for the maximum radius, we propose 
 a phase diagram delimiting four regions for the spreading and retraction dynamics of impacting 
drops.

\section{Drop retraction dynamics: Generic Features}
As the impact dynamics of liquid droplets on a solid surface happens usually in a few tens of 
milliseconds, we use a high-speed video system (1000 frames/second, Photonetics) to analyze the 
drop-impact events.  When necessary, we use an ultrahigh-speed system allowing to go up to 
120,000 frames/second (Phantom V7).  We study aqueous drops impacting on a solid surface; the 
surface we used is Parafilm, which provides us with a hydrophobic surface (receding contact 
angle for water $\thetaR \approx 80^\circ$). In addition, the surface has a low contact angle hysteresis with 
water, and allows us to obtain highly reproducible results. The liquids we used are 
different water-glycerol mixtures. Varying the glycerol concentration, we vary the liquid 
viscosity, keeping the liquid density and its surface tension almost 
constant. For the highest concentration of glycerol, the surface tension has decreased from $72$ (pure water) to $59$ $mNm^{-1}$, whereas the density has increased to $1150 kg/m^3$. The viscosity is varied between $1$ and $205 \;{\rm mPas}$. Viscosity, density and surface tension were measured before each impact experiment. Drops were produced using precision needles, and the initial radius of the drops $\RI$ have been systematically measured on the images ($1.1 <\RI<1.4$ mm). From the high-speed images such as the ones shown 
in Fig.\ref{fig:Rwater}, we follow the contact radius $R$ in 
time. This section summarizes the results of more than 
$80$ different drop impact experiments, each of 
which have been repeated at least two times. 

Two series of experiments were performed: first, letting the droplets fall from a fixed height, 
but increasing the viscosity, we increase the Ohnesorge number while keeping the inertial 
forces constant.  The second series of experiments is performed at 
fixed viscosity and upon increasing the height from which the droplets falls; the droplet turns 
out to be in free fall (as is verified in the experiment to within a few percent) and so the relation between fall 
height $h$ and impact velocity is simply $\VI=\sqrt{gh}$, with $g$ the gravitational 
acceleration. Increasing the impact velocity increases the Weber number, keeping the Ohnesorge 
number fixed, where the Weber number, $\We$, compares
the inertial forces to the capillary forces, $\We \equiv \rho R_{\rm I} \VI^2/\gamma$. 

In all that follows, we restrain ourselves to high-speed impact conditions. More precisely, the 
Weber and Reynolds numbers  are chosen so that 
$\We>10$ and $\Rey>10$, where 
$\Rey\equiv\rho\RI\VI/\eta$ is the 
Reynolds number. This implies that inertial forces are  
at least one order of magnitude larger 
than both the capillary and the viscous forces.  Such 
conditions imply large deformations of the drop 
when the liquid impinges on the solid substrate. 
On the other hand, we also restrain our 
experiments to impact speeds that are far from 
the 'splashing' regime in which the drop 
disintegrates after impact to form a collection of much smaller 
droplets \cite[]{MST95}.

The pictures in Fig.\ref{fig:Rwater} show that two distinctly 
different regimes exist for the shape of the 
droplets after impact. For low fluid viscosity, we typically obtain 
the images shown in Fig.\ref{fig:Rwater}(a). At the onset of retraction, almost all of the fluid is contained 
in a donut-shaped rim, with only a thin film of 
liquid in the center. On the other hand, for high viscosities the deformation of the 
drop is less important, and the pancake-shaped droplet of Fig.\ref{fig:Rwater}(b) results. These visual 
observations allow to distinguish the capillary-inertial and the capillary-viscous regimes that are described in detail below directly.

\subsection{Drop Retraction Rate: influence of fall height and viscosity}
\begin{figure}
\centerline{\includegraphics[width=15cm]{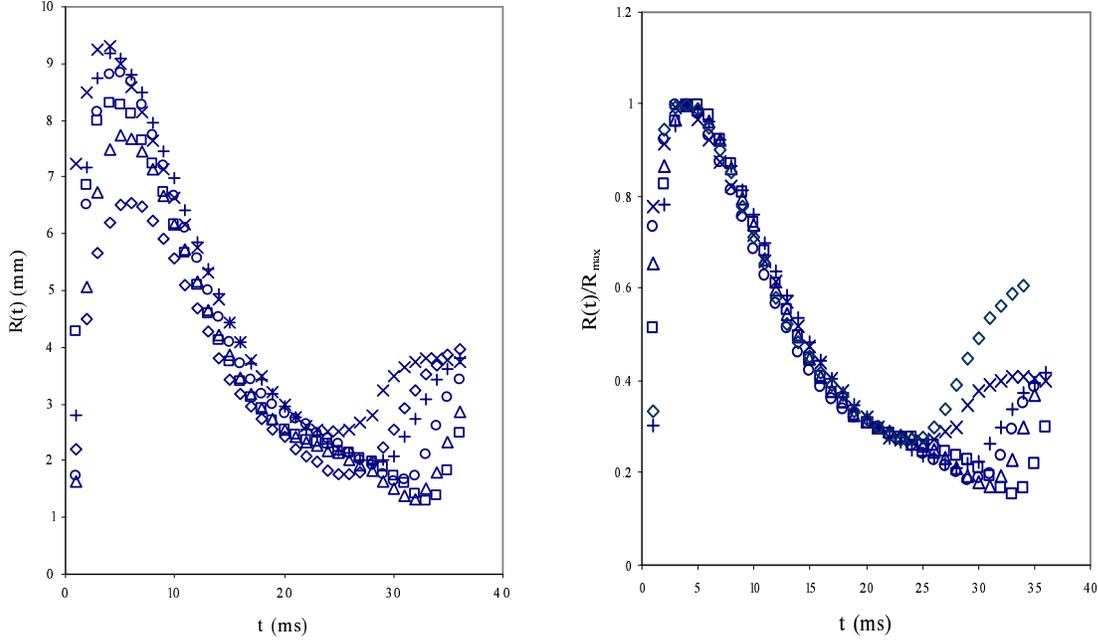}}
\caption{Temporal evolution of the contact radius 
for a water-glycerol drop $\Oh=9.1\,10^{-2}$, $\RI=1.2$ mm 
(a) contact radius vs. time, (b) contact radius normalized by the maximum spreading radius vs. time
Impact velocities : 
$\times$: $V_{I}=2.4{\rm ms}^{-1}$, 
$+$: $V_{\rm I}=2.2{\rm ms}^{-1}$, $\circ$: 
$V_{\rm I}=1.9{\rm ms}^{-1}$,$\square$: $V_{\rm I}=1.7{\rm ms}^{-1}$, $\Delta$: 
$V_{\rm I}=1.4{\rm ms}^{-1}$, $\diamond$: $V_{\rm I}=1{\rm ms}^{-1}$ }
\label{fig:Rmax}
\end{figure}

Fig. \ref{fig:Rmax} summarizes the most important findings of this study. The temporal evolution of the drop contact radius $R(t)$ for different impact velocities, shown in (a), is normalized in (b) by its 
maximal value at the end of the  spreading $\Rmax$. Two important observations are made. (i) A well defined retraction velocity $V_{\rm ret}$ can be extracted from each experiment; this is a 
non-trivial observation that will be rationalized 
below. (ii) Independently of the impact speed, all the $R(t)/\Rmax$ curves collapse onto a single curve
for different impact velocities. This shows that 
the retraction {\it rate}, rather than the 
retraction speed is the natural quantity to 
consider, and that this rate is independent of 
the impact velocity.  These results hold for all the viscosities tested in our experiments.

\begin{figure}
\centerline{\includegraphics[width=12cm]{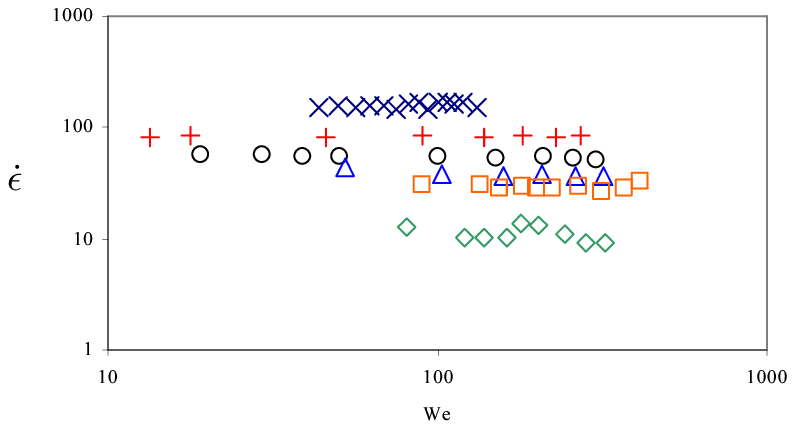}}
\caption{Retraction Rate plotted versus Impact 
Weber number for various water glycerol droplets. 
$\times$: $\Oh=2.510^{-3}$, $+$: 
$\Oh=3.910^{-3}$, $\circ$: $\Oh=1.510^{-2}$, 
$\vartriangle$: $\Oh=1.610^{-2}$, $\square$: 
$\Oh=2.310^{-2}$, $\diamond$: 
$\Oh=7.110^{-2}$}\label{fig:TR-We}
\end{figure}

In Fig. \ref{fig:TR-We} we have plotted the 
retraction rate $\dot \epsilon\equiv\Vret/\Rmax$ 
versus the impact Weber number, where $\Vret$ is 
defined by $\Vret\equiv\max{[-\dot R(t)]}$. 
Clearly, the drop retraction rate does not depend on the impact 
velocity. One might think that the explanation 
for this observation is rather obvious: the 
initial kinetic energy of the droplet is
transformed into surface energy (which fixes 
$\Rmax/R_I \propto \We^{1/2}$), and is then transformed back into kinetic 
energy (which in turn fixes $V_{\rm ret}\propto V_I$). This naive explanation is unfortunately wrong fro the following reasons.
First, it has been observed recently 
that, at the onset of retraction, low viscosity liquids 
undergo vortical motion in the drop \cite[]{Clanet2004}. This 
residual flow in the drop reveals that a part of 
the initial kinetic energy is still available then, and thus that a simple energy 
balance argument cannot work. This was indeed already suggested by previous observations of a clear disagreement
between experiments and the $\Rmax/R_I \propto \We^{1/2}$ law\cite[]{fukai93,Roisman2002,Okumura2003}. The second reason why the simple energy-balance argument does not work follows directly from Fig. 
 \ref{fig:TR-We}, where it is shown that 
the retraction rate depends on the 
viscosity and consequently that the previous inviscid picture 
is not correct.  

We therefore performed experiments that elucidate the role of the 
viscosity, or, equivalently, of the Ohnesorge 
number. For what follows, it is convenient to 
define two intrinsic time scales for the droplet: 
a viscous one and an inertial one. The viscous 
time is the relaxation time of a large-scale 
deformation of a viscous 
drop: $\tauv\equiv(\eta\RI)/\gamma$, whereas the 
inertial time scale: $\taui=(\frac{4}{3}\pi\rho\RI^3/\gamma)^{1/2}$ 
corresponds to the capillary oscillation period of a perturbed inviscid 
droplet. Since $\taui$ is 
independent of $\VI$ and $\eta$, this quantity is 
almost constant for all tested drops.

\begin{figure}
\centerline{\includegraphics[width=12cm]{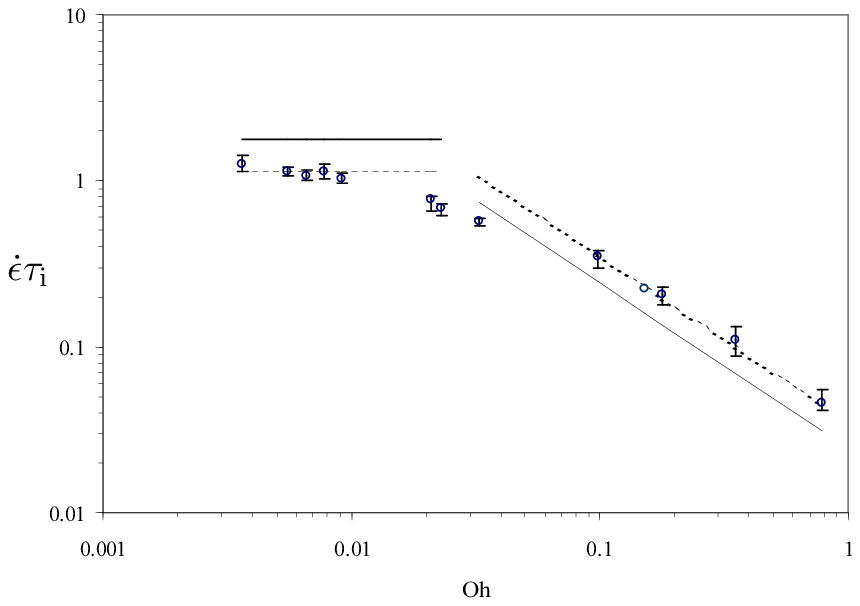}}
\caption{Circles: Normalized retraction rate $\dot{\epsilon}\taui$ plotted 
versus the Ohnesorge number, experimental values. Error bars represent the maximum deviation from the mean value.
Full line: (left) 
$\dot{\epsilon}\taui$ evaluated using Eq.
\ref{Eq:inertial}, (right) 
$\dot{\epsilon}\taui$ evaluated using 
Eq.\ \ref{eq:viscous}. Dashed line :(left) Fit obtained taking the mean value of the five first experimental points,(right) Best fit according to the predicted $1/\Oh$ power law.}\label{fig:TR-Oh}
\end{figure}

Fig. \ref{fig:TR-Oh} shows the retraction rate, 
made dimensionless using the inertial time, as a 
function of the Ohnesorge number. It can be observed in the figure that two different regimes exist 
for the retraction rate. The first 
region where the retraction rate $\dot{\epsilon}$ is independent of the 
viscosity points to an inertial regime and $\dot{\epsilon}\propto \taui^{-1}$. The 
retraction rate is consequently found not to depend on the impact 
speed, a result similar to that obtained recently by \cite[]{Richard2002} who show that the contact 
time is independent of the impact speed. For higher viscosities, typically $\Oh >0.05$, the retraction rate 
decreases strongly. In this regime, capillary and viscous forces govern the
dynamics: we find $\dot{\epsilon}\propto \tauv^{-1}$. 

\section{Two simple models for the drop retraction dynamics}
We have consequently established the existence of 
two different regimes for the retraction rate: a 
viscous one and an inertial one. We now develop  
some simple arguments allowing for a 
semi-quantitative description of the dynamics, 
using ideas already existing for the dynamics of 
dewetting, a problem closely related to the 
current one.

\subsection{Inertial regime}
We employ a 
Taylor-Culick approach commonly used for the
inertial dewetting of thin films \cite[]{Taylor,Culick,Buguin} to describe the drop retraction rate. For 
high-velocity
drop impacts, liquid spreads out into a thin film of 
thickness $h$ and radius $\Rmax$. The liquid
subsequently dewets rapidly the surface, and in doing so 
forms a rim that collects the liquid  that is initially
stored in the film. The shape of the drop surface shape is therefore never in a steady state and consists of
a liquid film formed during the spreading stage and a receding rim. The contact angle at the outer side of the rim is taken to be
very close to the receding contact angle ($\thetaR$) since viscous effects can be neglected small \cite[]{Buguin}.
The dynamics is consequently determined by a competition between capillary tension coming from the
thin film and the inertia of the rim.
If we write down momentum conservation for the liquid rim:
\begin{equation}
\frac{\rm d}{\rm d t}\left(m\frac{\rm dR(t)}{\rm 
d t}\right)=F_{\rm C}\label{eq:Culick}
\end{equation}
with $m$ the mass of the liquid rim and $F_{\rm 
C}$ the capillary force acting on it, $F_{\rm 
C}\sim 2\pi\gamma R(t)\left[1-\cos(\theta_{\rm 
R})\right]$. The stationary solution of 
Eq.\ref{eq:Culick} can be obtained writing 
$\dot{m}(t)=2\pi \rho R \Vret h$, and gives:
$\Vret=\sqrt{\gamma[1-\cos(\thetaR)]/(\rho h)}$. 
Using volume conservation, 
$h\sim\frac{4}{3}\RI^3\Rmax^{-2}$, it follows 
that:
\begin{equation}
\frac{\Vret}{\Rmax}\sim \taui^{-1}\sqrt{\pi\left[1-\cos\thetaR)\right]}
\label{Eq:inertial}
\end{equation}
Which is the final result. Comparing with the 
experimental data, it turns out that this 
equation not only gives the correct scaling 
behavior for the retraction in this regime rate but also 
provides a rather accurate estimate of the 
numerical prefactor (see Fig~\ref{fig:TR-Oh}). Indeed, the ratio between the experimental and the predicted numerical prefactors is found to be $0.6$ Repeating the experiment for water on a polycarbonate surface, which changes the contact angle value to $60 ^\circ$, we retrieve exactly the same ratio of $0.6$.  

\subsection{Viscous regime}
In the opposite limit of very viscous liquids, the drops adopt pancake shapes upon impact. During the 
first stages of retraction, the pancake shape rapidly relaxes towards a roughly 
spherical cap, and the drop shape remains like this during the retraction since the capillary number is small.  During the retraction, it is only the contact angle that varies slowly: it is mainly this slow contact angle dynamics that dictates the drop evolution during the retraction. Contrary to the previous analysis, the slow receding velocity allows to assume a quasi-static
dynamics for the surface shape during the retraction. 
In this regime, it is then natural to assume that the work done by the capillary force 
$F_{\rm C}$ is dissipated 
through viscous flow near the contact line. Since we focus our study on high-speed impacts, $\Rmax$ is always much larger that $\RI$ which justifies a small $\theta(t)$ approximation at the onset of retraction. The viscous effects near the contact line 
then lead to the well-known linear force-velocity relation~\cite[]{PGG85}:
\begin{equation}
F_{\rm V}=-\frac{6\pi\eta}{
\theta}\ln\left(\frac{\Lambda}{\lambda}\right)R(t)\dot{R(t)}\label{forcevitesse}
\end{equation}
where $\Lambda$ and $\lambda$ are respectively a 
macroscopic and a microscopic cutoff lengths. $\Lambda$ is typically of 
the same order as the drop size $\sim1{\rm mm}$. 
$\lambda$ is a microscopic length, and is usually 
taken to be on the order of $\lambda\sim 1{\rm 
nm}$~\cite[]{PGG85}.
On the other hand, the capillary force drives the 
retraction. Near the contact line it can be written:
\begin{equation}
F_{\rm C}=2\pi R(t) \gamma \left [\cos 
\theta(t)-\cos 
\thetaR\right]\label{forcecappilaire}
\end{equation}
Volume conservation gives: $\frac{4}{3}\pi 
\RI^3\sim\frac{\pi}{4}\theta(t)R^3(t)$, where we 
have taken the small angle limit. Eqs. 
\ref{forcevitesse} and \ref{forcecappilaire} together 
with the volume constraint leads to the following 
relation for the variation of the contact radius:
\begin{equation}
\frac{\dot{R}(t)}{R(t)}=-\frac{\left[1-\frac{1}{2}\theta^2(t)-\cos(\theta_R)\right]\theta(t)^{4/3}}{(144)^{1/3} \ln(\Lambda/\lambda)}\tauv^{-1}
\label{eq:theta}
\end{equation}
the above equation is obtained in the small angle 
limit and is only valid for short time after the 
onset of retraction. We estimate the retraction rate $\dot{\epsilon}$ as the maximum value of ${\dot{R}(t)}/{R(t)}$ so that: 
\begin{equation}
\frac{\Vret}{\Rmax}\approx\left( \frac{3}{25} \right)^{1/3} \frac{(1-\cos \thetaR)^{5/3}}{5 
\ln(\Lambda/\lambda)}\tauv^{-1} \label{eq:viscous}
\end{equation}
Comparing again to the experiments, good 
agreement is found: the retraction rate is solely 
set by the viscous relaxation time $\tauv$ and consequently 
$\dot{\epsilon}\taui \propto \Oh^{-1}$. Beyond this 
correct scaling prediction, Eq. \ref{eq:viscous} 
provides a quite accurate estimate for the 
numerical prefactor as is shown in 
Fig~\ref{fig:TR-Oh}. Indeed, the ratio between the experimental and the predicted numerical prefactors is found to be $1.5$. Again, repeating the experiment on a polycarbonate surface, this ratio changes only slightly from $1.5$ to $1.8$.

\section{Conclusions and perspectives}
\begin{figure}
\centerline{\includegraphics[width=14cm]{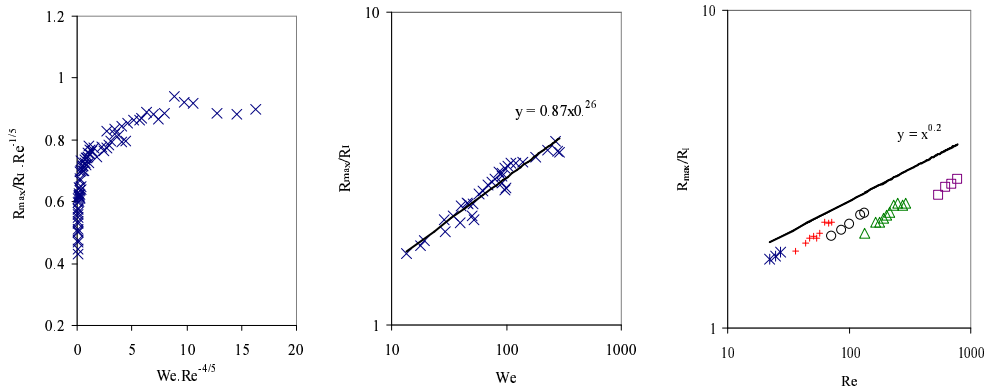}}
\caption{(a) Normalized maximum spreading radius plotted vs. the impact number. (b) $\Rmax$ (normalized by the radius before impact) plotted vs. Weber number for small values of the impact number. Full line: power-law fit.  (c) $\Rmax$ (normalized by the radius before impact) plotted vs. Reynolds number for large values of the impact number. Full line: predicted power-law dependence with power $0.2$. 
$*$: $\eta=10^{-1}Pa.s$, 
$+$: $9.510^{-2}$, 
$\circ$: $\eta=4.810^{-2}Pa.s$, 
$\vartriangle$: $\eta=2.810^{-2}Pa.s$, $\square$: $\eta=10^{-2}Pa.s$}
\label{fig:RmaxWe}
\end{figure}
Our experiments reveal that the retraction rate is  independent of the impact speed. To account for the retraction speed, the maximum radius to which the droplet 
expands, has to be known also.
A number of studies have been devoted to the understanding of the maximum 
spreading radius (see for instance \cite[]{fukai93,Roisman2002,Clanet2004}). However, no clear 
and unified picture emerges from previous 
experimental investigations. A recent experimental study of $\Rmax$, combined with  recent theoretical ideas in the same spirit of the ones presented here was done by \cite{Clanet2004}. They obtain a zeroth order (asymptotic) description of the 
spreading stage, compare it with experiments and
suggest that two asymptotic regimes exist for $\Rmax$. The first is given by a 
subtle competition between the inertia of the droplet  and the capillary forces; if only 
these two are important, it follows that $\Rmax/\RI\propto\We^{1/4}$. In the 
second regime, $\Rmax$ is given by a balance between inertia and viscous 
dissipation in the expanding droplet, leading to $\Rmax/\RI\propto\Rey^{1/5}$. Consequently, 
 a single dimensionless number is  defined that discriminates 
between the two regimes: $\I=\We\Rey^{-4/5}$ referred to as the Impact number. 
The crossover between the two regimes happens at a $P$ of order unity. 

Our 
experimental data are in qualitative agreement with their prediction, as is 
shown in Fig. \ref{fig:RmaxWe}.a. At low $\I$, the scaling $R_{\rm max}/\RI\sim\We^{1/4}$ is 
clearly observed. However, for impacts corresponding to $\I>1$, we observe only a very 
slow variation of the maximum spreading radius as a function of $\I$. Therefore, 
the relation between  $\Rmax$ and the 
Reynolds number is not very clear from our data (Fig \ref{fig:RmaxWe}. c).  Although the main trend is not in 
strong contradiction with the prediction $\Rmax/\RI\propto\Rey^{1/5}$, a power-law fit of our data
gives exponents that are always smaller than the predicted value of $0.2$. Perhaps even more 
important- in view of the small range of the maximal expansion $\Rmax$ that we cover- is that the 
different water-glycerol mixtures do not appear to collapse on a single master 
curve, as would be predicted by the above argument. However, 
since the maximum value of $\I$ that we reach is on the order of 10, it may be 
that we have not reached the purely viscous regime. In that case, the capillary, 
inertial and viscous forces are still of comparable amplitude and have to be taken into account together. Note also that the more sophisticated 
models reviewed in \cite{Ukiwe2004} do not provide better agreement with our experimental 
measurements.  

\begin{figure}
\centerline{\includegraphics[width=10cm]{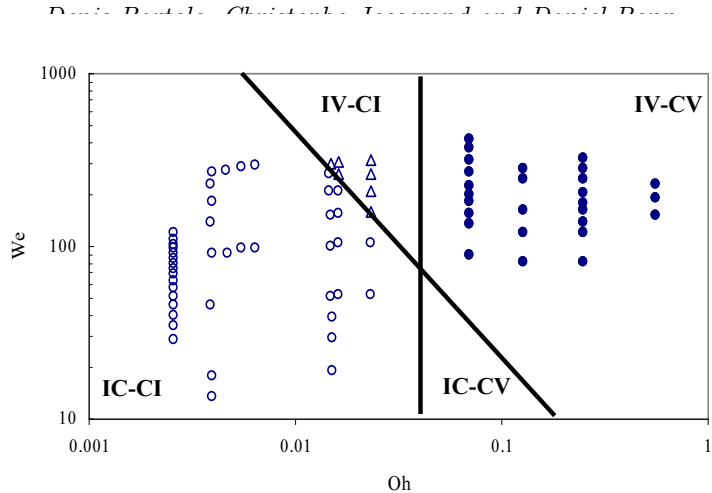}}
\caption{Phase diagram in the $(\We,\Oh)$ plane for the impact and retraction dynamics of droplets. The four regions are discussed in the text, and the symbols represent the parameters of the data reported in this paper. Different symbols have been assigned for each region.}\label{fig:WeOh}
\end{figure}

Despite this small problem, we are now able to develop a simple unified picture for drop impact dynamics accounting
for both the spreading and the retraction dynamics. The 
two natural dimensionless numbers that have been identified are the impact 
number $\I$, that quantifies the spreading out of the droplet, and the Ohnesorge number $\Oh$ that 
quantifies the retraction. We can thus construct a phase diagram in the 
experimentally explored $(\Oh,\We)$ plane, which is shown on Fig. 
\ref{fig:WeOh}. The experimentally accessible plane is divided in four 
parts, where the main mechanisms at work during the impact process are different. These four parts are separated by the curves $\Oh=0.05$ and $\We=\Oh^{-4/3}$.
They are labeled as follows: ICCI the drop dynamics is given by a competition 
between inertia and capillarity both for the spreading and the retraction. IVCV: 
inertia and viscous forces dominate the spreading, capillary and viscous 
forces dominate the retraction. These two regimes have been studied in detail 
here. The two more intriguing regions are the IVCI (viscous spreading, inertial retraction) and 
ICCV (capillary spreading, viscous retraction) that are 
unfortunately  difficult to explore in detail. For the IVCI- regime, the large inertia at impact, 
combined with a small surface tension,
will make the droplets undergo large non-axisymetric deformations and they will eventually splash and disintegrate. 
On the other end of the phase diagram, the 
ICCV region corresponds to very low impact speeds and important capillary forces, implying very small deformations of the droplets. If the 
deformations are small, pinning of the contact line of the droplets will become important, 
and all our simple scaling arguments for both the maximum radius and the 
retraction rate are invalidated. 

A numerical investigation of droplet impact 
would be very helpful for two reasons. First, numerics would allow to vary 
$\RI$ while keeping  all the other physical parameters constant. This would 
allow to check the robustness of our results, since experimentally it is not 
easy to vary $\RI$ over a wide range. Second, as emphasized above, the viscous 
regime for the maximum radius is  difficult to characterize precisely due to the 
smallness of the variation of $\Rmax$ for viscous drops. If precise numerical 
simulations could be done, these different remaining problems could be 
resolved.

In sum, we have studied the retraction dynamics of liquid 
droplets upon high-speed impact on non-wetting solid 
surfaces. Perhaps the strongest conclusion from 
our investigation is that the rate of retraction 
of the droplet is a drop constant which does 
not depend on the impact velocity. Two regimes 
for the retraction rate have been identified: a 
viscous regime and an inertial regime. We have in 
addition shown here that simple hydrodynamic 
arguments can be formulated that give very 
reasonable agreement with experiments in the two 
different regimes.
{\bf Acknowledgments:} Benjamin Helnann-Moussa is acknowledged for help with the experiments.
Denis Bartolo is indebted to the CNRS for providing a post-doctoral fellowship. 
LPS de l'ENS is UMR 8550 of the CNRS, associated with the universities Paris 6 
and Paris 7.

\bibliographystyle{jfm}
\bibliography{Drop,biblio}

\end{document}